\documentstyle[12pt,ppeuc]{article}
 
\begin{document}
\newcommand{\newc}{\newcommand}
 
\newc{\be}{\begin{equation}}
\newc{\ee}{\end{equation}}
\newc{\ba}{\begin{eqnarray}}
\newc{\ea}{\end{eqnarray}}
\newc{\bea}{\begin{eqnarray}}
\newc{\eea}{\end{eqnarray}}
\newc{\D}{\partial}
\newc{\ie}{{\it i.e.} }
\newc{\eg}{{\it e.g.} }
\newc{\etc}{{\it etc.} }
\newc{\etal}{{\it et al.} } 
\newc{\ra}{\rightarrow}
\newc{\lra}{\leftrightarrow}
\newc{\no}{Nielsen-Olesen }
\newc{\lsim}{\buildrel{<}\over{\sim}}
\newc{\gsim}{\buildrel{>}\over{\sim}}
 
May 1997\hfill       
CRETE-97/15 ,  DEMO-HEP-97/10 \\
    
\vskip 0.5cm
 
\title{
Domain Walls and Vortices with Non-Symmetric Core
\footnote{ Talk presented at the "Particle Physics and the Early Universe"
Conference, April 7-11, Univ. of Cambridge.}  
}
\author{\addlink {Minos Axenides}{mailto:
 axenides@gr3801.nrcps.ariadne-t.gr}}
\institute{Institute of Nuclear Physics,\\ N.C.S.R. Demokritos \\
153 10, Athens, Greece 
}
\vspace{0.5cm}
\author{\addlink {Leandros Perivolaropoulos}{
http://www.edu.physics.uch.gr/~leandros}}
\institute{Department of Physics,\\ University of Crete \\
71 003 Heraklion, Greece 
} 
\vskip .1in
\begin{abstract}  
\noindent 
We review recent work on a new class of  topological defects 
which possess a nonsymmetric core.
They arise in scalar field theories with global symmetries, 
U(1) for domain walls and SU(2) for vortices, which are explicitly 
broken to $Z_2$ and U(1) respectively.  
Both of the latter symmetries are spontaneously broken. 
For a particular range of parameters both types of defect solutions are 
shown to become unstable and decay to the well known
stable walls and vortices with symmetric cores.
\end{abstract} 

\section{Introduction}

Topological defects[\cite{vs94,v85}] are stable field configurations 
(solitons[\cite{r87}]) that arise in 
field theories with spontaneously broken discrete or continous symmetries.
Depending on the topology of the vacuum manifold $M$ they are usually 
identified
as domain walls[\cite{v85}] (kink solutions[\cite{r87}]) when
$M=Z_2$, as strings[\cite{no73}] and one-dimensional textures 
(ribbons[\cite{bt94,bt95}]) when $M=S^1$, as monopoles 
(gauged[\cite{t74,p74,dt80}]
or global[\cite{bv89,p92c}]) and two dimensional textures 
($O(3)$ solitons [\cite{bp75,r87}])when $M=S^2$ and three dimensional 
textures [\cite{t89}]
(skyrmions[\cite{s61}]) when $M=S^3$. 
They are expected to be remnants of phase transitions [\cite{k76}] that may
have occured in the early universe. They also form in various condenced matter
systems which undergo low temperature transitions [\cite{z85}].
Topological defects appear to fall in two broad categories.   
In the first
one the topological charge becomes non-trivial due to the behavior of the
field configuration at spatial infinity. The symmetry of the vacuum
gets restored at the core of the defect.
Domain
walls, strings and monopoles belong to this class of {\it symmetric defects}.

In the second category the vacuum manifold gets covered completely as the field
varies over the whole of coordinate space.
Moreover its value at infinity is identified
with a single point of the vacuum manifold. Textures [\cite{t89}] 
(skyrmions [\cite
{s61}]), $O(3)$ solitons [\cite{bp75}] 
(two dimensional textures [\cite{t89}]) 
and ribbons 
[\cite{bt94}] belong to this class which we
will call for definitenss {\it texture-like} defects. 
The objective of the present discussion is to present examples of defects 
which belong
to neither of the two categories, namely 
the field variable covers the whole vacuum manifold at infinity with the core
remaining in the non-symmetric phase. For definiteness we will call 
these {\it non-symmetric} defects.

Examples of nonsymmetric defects have been discussed previously in
the literature. Everett and Vilenkin [\cite{ve82}],in particular, pointed out 
the existence of domain
walls and strings with non-symmetric cores which are unstable though
to shrinking and collapse due to their string tension. 
A particular case of non-symmetric {\it gauge} defect was recently considered by
Benson and Bucher [\cite{bb93}]  
who pointed out that the
decay of an electroweak semilocal string leads to a gauged "skyrmion"
with non-symmetric
core and topological charge at infinity. This skyrmion however, rapidly 
expands and
decays to the vacuum.
 
\par
In the present talk we review recent work where
we presented more examples of topological defects that
belong to what we defined as the "non-symmetric" class. We will study in
detail the properties of global domain
walls in section $2$  and of global vortices in section $3$.
In both cases we will identify the
parameter ranges for stability of the configurations with either
a symmetric or a non-symmetric core. For the case of a domain wall wall
we will discuss results of a simulation for an expanding bubble of a domain
wall. 

Finally, in section 4 we conclude, summarize and discuss the outlook of this
work.

\section{Domain Walls with NonSymmetric Core}

We consider a model with a $U(1)$ symmetry explicitly broken to a $Z_2$. This
breaking can be realized by the Lagrangian density [\cite{vs94,ap97}] 
\begin{equation}
{\cal {L}}={1\over 2}\partial _\mu \Phi ^{*}\partial ^\mu \Phi +
{Ì^2 \over 2}|\Phi |^2 + {m^2 \over 2}Re(\Phi ^2) - {h \over 4}|\Phi |^4
\end{equation}
where $\Phi =\Phi _1+i\Phi _2$ is a complex scalar field. After a rescaling 
\begin{eqnarray}
\Phi &\rightarrow &{m\over \sqrt{h}} \Phi \\
x  & \rightarrow & {1\over m} x \\
M & \rightarrow & \alpha m 
\end{eqnarray}

\par
The corresponding equation of motion for the field $\Phi$
is
\be
{\ddot \Phi} - {\nabla ^2} \Phi - (\alpha^2 \Phi + \Phi^*) +
|\Phi|^2 \Phi = 0
\ee

The potential takes the form
\be
V(\Phi) = -{m^4 \over {2
h}} (\alpha^2 |\Phi|^2 + Re(\Phi^2) - {1\over 2} |\Phi|^4)
\ee
For $\alpha
< 1$ it has the shape of a "saddle hat" potential 
i.e. at $\Phi = 0$ there is a local
minimum in the $\Phi_2$ direction but a local maximum in the $\Phi_1$ (Fig
1). For this range of values of 
$\alpha$ the equation of motion admits the well known  static kink solution
\bea
\Phi_1
&=& \Phi_R \equiv \pm (\alpha^2 + 1)^{1/2} \tanh (({{\alpha^2 + 1}\over 2})^{1/2}
x) \\
\Phi_2 &=& 0
\eea
It corresponds to a {\it symmetric} domain
wall since in the core of the soliton the full symmetry of the
Lagrangian is manifest ($\Phi(0) = 0$) and the
topological charge arises as a consequence of the behavior of the 
field at infinity 
($Q={1 \over 2}(\Phi(-\infty) - \Phi(+\infty))/ (\alpha^2 + 1)^{1/2} $).

For $\alpha > 1$ the local minimum in the $\Phi_2$ direction becomes a
local maximum but the vacuum manifold remains disconnected, and the $Z_2$
symmetry remains. This type of potential may be called a "Napoleon hat" 
potential in
analogy to the Mexican hat potential that is obtained in the limit $\alpha
\rightarrow \infty$ and corresponds to the restoration of the $S^1$ vacuum
manifold.

\begin{figure}
\begin{center}
\unitlength1cm
\begin{picture}(6,3)
\put(-3.5,-7.0){\includegraphics{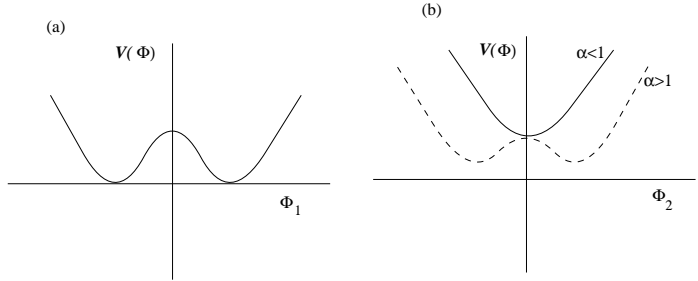}}
\end{picture}
\end{center}
\caption{(a) The domain wall potential has a local maximum
at $\Phi = 0$ in the $\Phi_1$ direction. (b) For $\alpha > 1$ ($\alpha < 1$)
this point
is a local maximum (minimum) in the $\Phi_2$ direction.
}
\end{figure}

The form of the
potential however implies that the symmetric wall solution may not be stable
for $\alpha >1$ since in that case the potential energy favors a solution
with $\Phi_2 \neq 0$. However, the answer is not obvious because for $\alpha
> 1$, $\Phi_2 \neq 0$ would save the wall some potential energy but would cost
additional gradient energy as $\Phi_2$ varies from a constant value at $x=0$
to 0 at infinity. Indeed a stability analysis was performed by 
introducing a small
perturbation about the kink solution reveals the presence of negative modes
for $\alpha > \alpha_{crit}= \sqrt{3} \simeq 1.73$
 For the range of values $ 1 < \alpha < 1.73 $ the potential takes the shape
of a "High Napoleon hat".   
We study the full non-linear static
field equations obtained from (6) for a typical value of $\alpha=1.65$
  with boundary conditions
\bea
\Phi_1 (0)
&=& 0 \hspace{1cm} \lim_{x\rightarrow \infty} \Phi_1 (x) = (\alpha^2
+1)^{1/2} \\
{\Phi_2^\prime} (0) &=& 0 \hspace{1cm} \lim_{x\rightarrow
\infty} \Phi_2 (x) = 0
\eea

\begin{figure}
\begin{center}
\unitlength1cm
\begin{picture}(10,4)
\put(-3.0,-1.7){\includegraphics{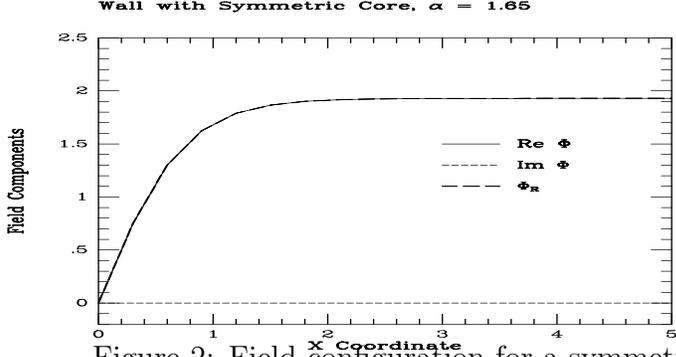}}
\end{picture}
\end{center}
\caption{Field configuration for a symmetric wall with $\alpha = 1.65$.
}
\end{figure}

Using a relaxation method based on collocation
at gaussian points [\cite{numrec}] to solve the system (6) 
of second order non-linear
equations we find that for $ 1 <\alpha < \sqrt{3}$ the solution relaxes to the
expected form of (7) for $\Phi_1$ while $\Phi_2 = 0$ (Fig. 2). For $\alpha > 
\sqrt{3}$ we find $\Phi_1 \neq 0$ and $\Phi_2 \neq 0$ (Fig. 3) obeying the
boundary conditions (13), (14) and giving the explicit solution for the
non-symmetric domain wall. In both cases we also plot the analytic solution
(7) stable only for $\alpha < \sqrt{3}$ for comparison (bold dashed line). As
expected the numerical and analytic solutions are identical for $\alpha < 
\sqrt{3}$ (Fig. 2).

\begin{figure}
\begin{center}
\unitlength1cm
\begin{picture}(10,4)
\put(-3.0,-1.7){\includegraphics{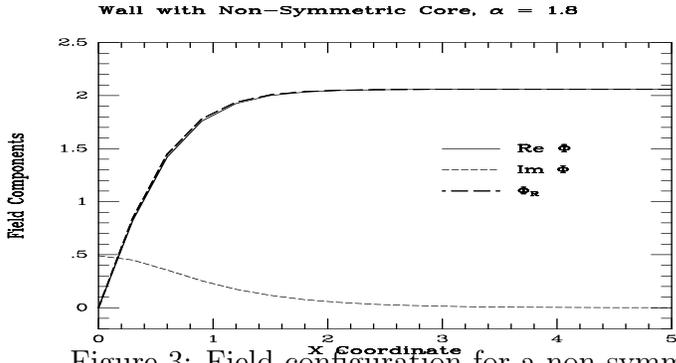}}
\end{picture}
\end{center}
\caption{Field configuration for a non-symmetric wall with $\alpha = 1.8$.
}
\end{figure}

\par
We now proceed to present results of our study on the 
evolution of bubbles of a domain wall. We  
constructed a two dimensional simulation of the field evolution of domain
wall bubbles with both symmetric and non-symmetric core. In particular we
solved the non-static field equation (6) using a leapfrog algorithm 
[\cite{numrec}]
with reflective boundary conditions. We used an $80 \times 80$ lattice 
and in all
runs we retained ${{dt} \over {dx}} \simeq {1\over 3}$ thus satisfying the
Cauchy stability criterion for the timestep $dt$ and the lattice spacing $dx$%
. The initial conditions were those corresponding to a spherically symmetric
bubble with initial field ansatz
\be
\Phi (t_i) = (\alpha^2 + 1)^{1/2}
\tanh [({{\alpha^2 +1} \over 2})^{1/2} (\rho - \rho_0)] + 
i \hspace{2mm} 0.1 \hspace{2mm}
e^{- ||x| - \rho_0|} {x \over {|x|}}
\ee
where $\rho = x^2 + y^2$ and $%
\rho_0$ is the initial radius of the bubble. Energy was conserved to within
2\% in all runs. For $\alpha$ in the region of symmetric core stability the
imaginary initial fluctuation of the field $\Phi (t_i)$ decreased and the
bubble collapsed due to tension in a spherically symmetric way as
expected.

\begin{figure}
\begin{center}
\unitlength1cm
\begin{picture}(6,8)
\put(-3.0,-2.0){\includegraphics{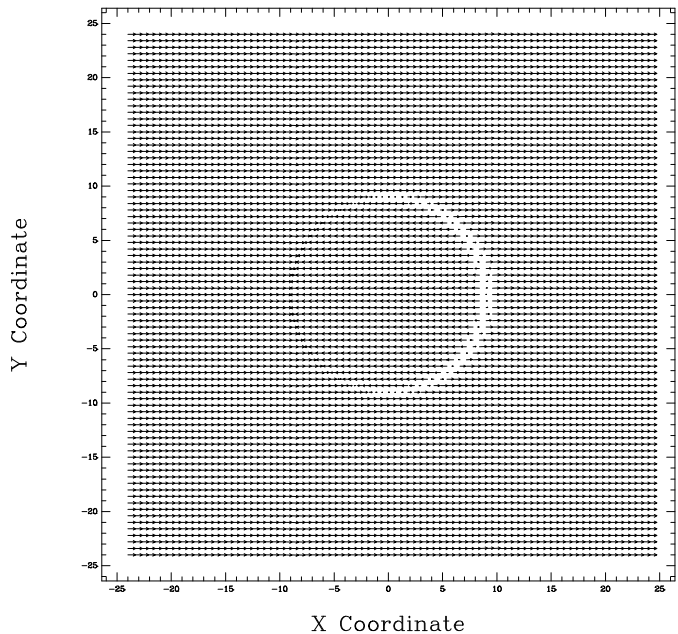}}
\end{picture}
\end{center}
\caption{Initial field configuration for a non-symmetric 
spherical bubble wall with $\alpha = 3.5$.
}
\end{figure}

For $\alpha$ in the region of values corresponding 
to having a non-symmetric stable core the
evolution of the bubble was quite different. The initial imaginary
perturbation increased but even though dynamics favored the increase of the
perturbation, topology forced the $Im\Phi (t)$ to stay at zero along a line
on the bubble: the intersections of the bubble wall with the y axis
(Figs. 4, 5). Thus in the region of these points, surface energy (tension)
of the bubble wall remained larger than the energy on other points of the
bubble. The result was a non-spherical collapse with the x-direction of the
bubble collapsing first (Fig. 5). 

\begin{figure}
\begin{center}
\unitlength1cm
\begin{picture}(6,8)
\put(-3.0,-2.0){\includegraphics{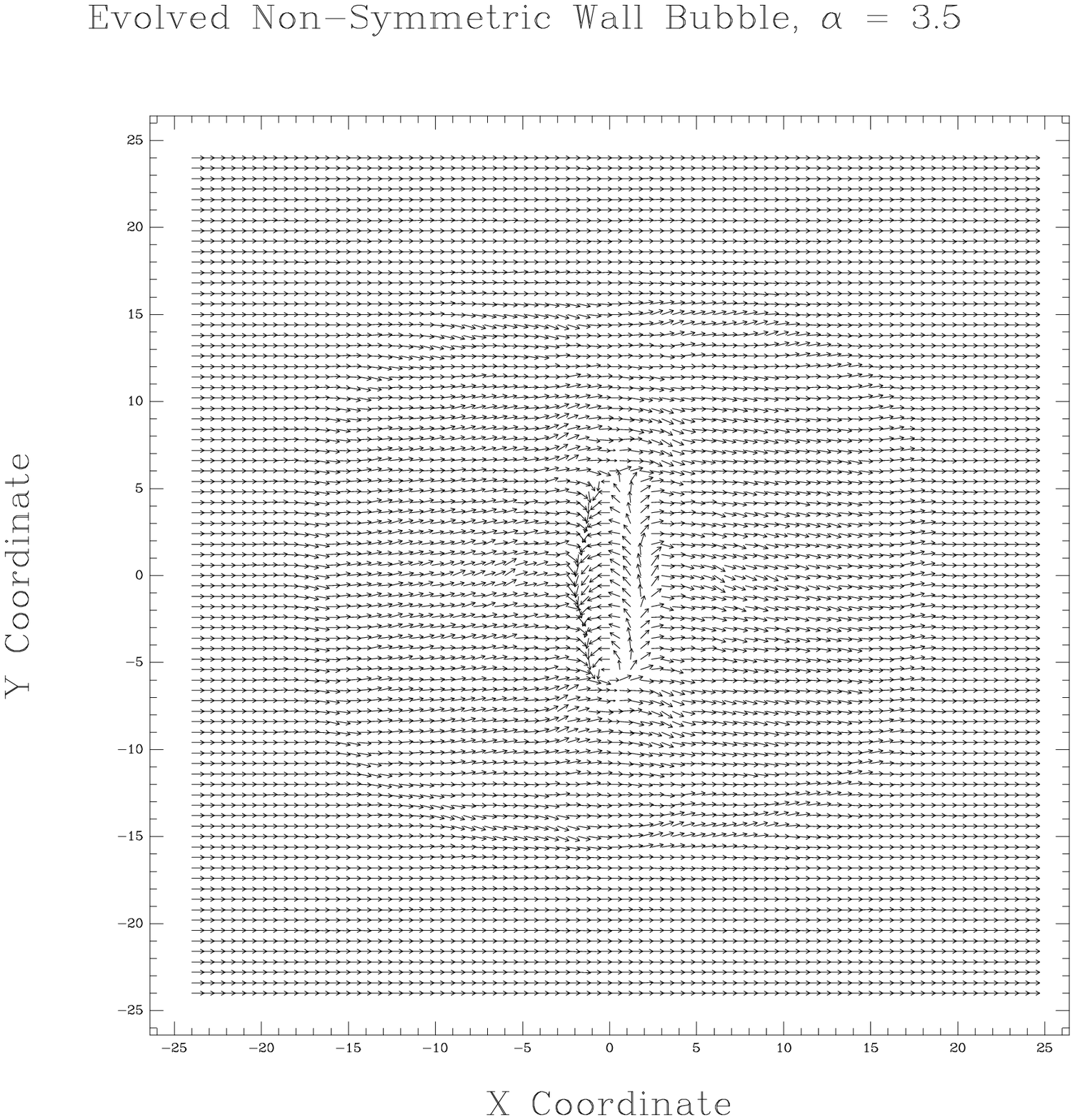}}
\end{picture}
\end{center}
\caption{Evolved field configuration ($t=14.25$, 90 timesteps) for a non-symmetric 
initially spherical bubble wall with $\alpha = 3.5$.
}
\end{figure}

\section{Vortices with Nonsymmetric Core}

We have generalized our analysis for domain walls to the case
of a scalar field theory that admits global vortices. We consider a model 
with an $%
SU(2)$ symmetry explicitly broken to $U(1)$. Such a theory is described by
the Lagrangian density:
\be
{\cal{L}} = {1 \over 2} \D_\mu {\Phi^\dagger}
\D^\mu \Phi + {M^2 \over 2} {\Phi^\dagger}\Phi + {m^2 \over 2} {\Phi^\dagger}
\tau_3 \Phi - {h\over 4} ({\Phi^\dagger}\Phi)^2
\ee
where $\Phi =
(\Phi_1, \Phi_2)$ is a complex scalar doublet and $\tau_3$ is the $2 \times 2$
Pauli matrix. After rescaling as in
equations (2)-(4) we obtain the equations of motion for $\Phi_{1,2}$
vspace{2cm}
\begin{figure}
\begin{center}
\unitlength1cm
\begin{picture}(6,8)
\put(-3.0,-1.7){\includegraphics{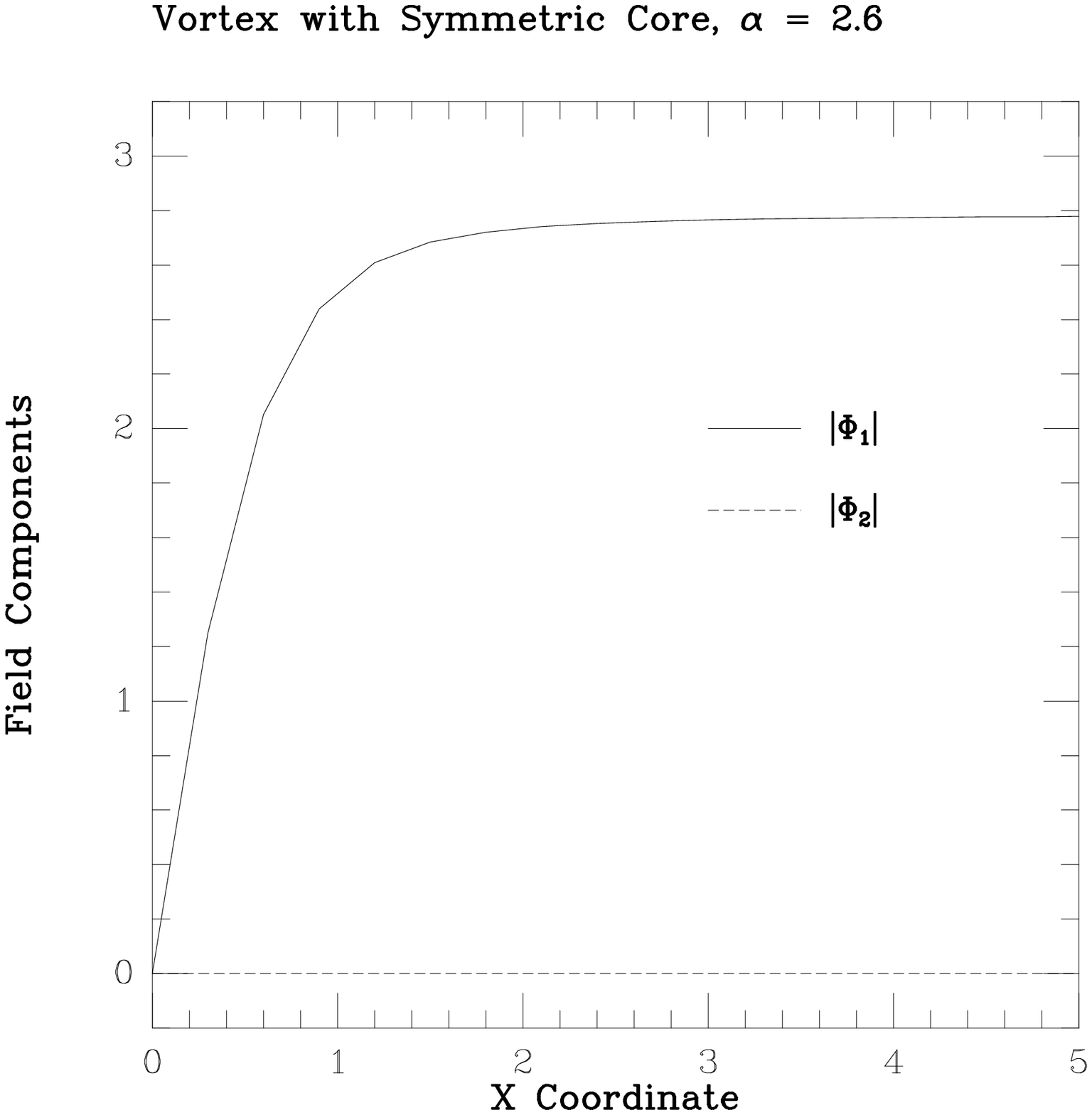}}
\end{picture}
\end{center}
\caption{Field configuration for a {\it symmetric-core}
global string with $\alpha = 2.6$.
}
\end{figure}

\be
\partial_\mu \partial^\mu \Phi_{1,2} - (\alpha^2 \pm 1) \Phi_{1,2}
+({\Phi^\dagger} \Phi) \Phi_{1,2} = 0
\ee
where the + (-) corresponds to
the field $\Phi_1$ ($\Phi_2$).
\par
Consider now the ansatz
\be
\Phi =
\left( \begin{array}{c}
\Phi_1 \\
\Phi_2
\end{array} \right)=
\left( \begin{array}{c}
f(\rho) e^{i\theta} \\
g(\rho)
\end{array}
\right)
\ee
with boundary conditions
\bea
\lim_{\rho \rightarrow 0} f(\rho)
&=& 0, \hspace{3cm} \lim_{\rho \rightarrow 0} { g^\prime} (\rho) = 0
\\
\lim_{\rho \rightarrow \infty} f(\rho) &=& (\alpha^2 +1)^{1/2},
\hspace{1cm} \lim_{\rho \rightarrow \infty} g (\rho) = 0 
\eea

\par

This ansatz
corresponds to a global vortex configuration with a core that can be either
in the symmetric or in the non-symmetric phase of the theory. Whether the
core will be symmetric or non-symmetric is determined by the dynamics of the
field equations. As in the wall case the numerical solution
of the system ($21$) of non-linear complex field equations with the
ansatz $(22)$ for various values of the parameter $\alpha$ 
reveals the existence of an $\alpha_{cr} \simeq 2.7 $
For $\alpha < \alpha_{cr} \simeq 2.7$ the
solution relaxed to a lowest energy configuration with $g(\rho) = 0$
everywhere corresponding to a vortex with symmetric core (Fig. 6).

For $
\alpha > \alpha_{cr} \simeq 2.7$ the solution relaxed to a configuration
with $g(0) \neq 0$ indicating a vortex with non-symmetric core (Fig. 7). Both
configurations are dynamically and topologically stable and consist
additional paradigms of the defect classification discussed in the
introduction. 

\begin{figure}
\begin{center}
\unitlength1cm
\begin{picture}(6,6)
\put(-3.0,-1.7){\includegraphics{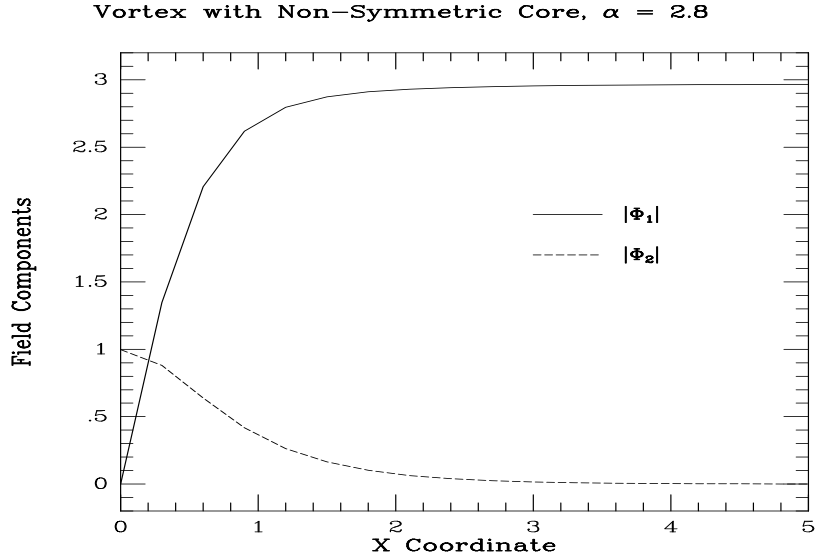}}
\end{picture}
\end{center}
\caption{Field configuration for a {\it non-symmetric-core}
global string with $\alpha = 2.8$.
}
\end{figure}
\par

\section{Conclusion}

We have studied the existence and stability properties of 
defects (domain walls and vortices) 
with non-symmetric core and non-trivial winding at infinity.
These defects arise in scalar field theories that exhibit an
explicit breaking of a global symmetry ,$ U(1)$ for domain wall
and $SU(2)$ for vortices. In their spectrum and for a particular
range of parameters
topologically stable and unstable defects appear with either symmetric

or nonsymmetric cores.
Possible implications for the cosmology of the early universe are the
following: 
With regard to the case of domain walls with a symmetric core (saddle and
high Napoleon hat potentials) a possible embedding of such configurations
in a realistic $2$Higgs electroweak model may realize a new mechanism
for baryogenesis at the electroweak phase transition. Defect mediated
baryogenesis has been so far only succesfully implemented at scales introduced
near or above the electroweak one \cite{bdpt96}.
 These mechanisms are based on unsuppressed
B+L violating sphaleron transitions taking place in the 
symmetric core of the defects during scattering processes
\cite{ckn93}.
As a result of our work the question of existence of electroweak domain walls 
with a symmetric core
 now translates to whether in the most general electroweak lagrangian
with two Higgs doublets potential energies
of the "saddle hat" or "high napoleon hat" type exist for an appropriate
range of parameters. As the parity symmetry in these models is broken
both spontaneously an explicitly the expected domain walls in their spectrum
are certainly of the nontopological type. 
Moreover it would be of interest to see if succh defects arise at a second
order electroweak phase transition.    
Our observation of non-spherical collapse  of 
wall bubbles with non-symmetric core may imply that the domain wall
network simulations need to be re-examined for parameter
ranges where a non-symmetric core in energetically favored. 

\par
With regard to our demonstration of existence of vortices with 
nonsymmetric core it becomes immediately suggestive the existence of a new
kind of a bosonic superconducting string , possesing massive charge carriers
[\cite{w85}].
This would be the case if our model is properly coupled to a $U(1)$
gauge field.The physics of fermions introduced to such a system is also  
open for investigation. The astrophysical and cosmological role of 
superconducting strings has been extensively investigated in the literature
[\cite{otw86,ds88}].

\section{Acknowledgements}
This work was supported by the E.U. grants $CHRX-CT93-0340$,  
$CHRX-CT94-0621$ and $CHRX-CT94-0423$ 
as well as by the Greek General Secretariat of Research and Technology grants 
$95E\Delta 1759$ and $\Pi E N E \Delta 1170/95$. 
We are particular thankful to D.A.M.T.P. of the
U.of Cambridge and Anne Davis for their hospitality and during our
stay at Cambridge.

\end{document}